\documentclass{elsart}
\usepackage{amsfonts}
\usepackage{graphicx}
\def\mg{^{28}\mathop{\mathrm{Mg}}\nolimits}
\def\pb{^{208}\mathop{\mathrm{Pb}}\nolimits}
\begin{document}
\begin{frontmatter}
\title{Nonlinear waves of nuclear density}
\author[jinr,IThF]{V.G. Kartavenko\thanksref{corr}},
\author[RomAc]{A. S\~andules\~cu} and
\author[IThF]{W. Greiner}
\address[jinr]{Bogoliubov Laboratory of Theoretical Physics,
Joint Institute for Nuclear Research, Dubna, Moscow District, 141980,
Russia}
\address[RomAc]{Romanian Academy, Calea Victoriei 125, Bucharest, 71102,
Romania}
\address[IThF]{Institut f\"ur Theoretische Physik der J. W. Goethe
Universit\"at\\
D-60054 Frankfurt am Main, Germany}
\thanks[corr]{Corresponding author. Phone: +7 09621 63710;
Fax: +7 09621 65084;\\
\hspace*{43mm}E-mail: kart@thsun1.jinr.ru}

\begin{abstract}
Nonlinear excitations of nuclear density are considered
in the framework of semiclassical nonlinear nuclear hydrodynamics.
Possible types of stationary nonlinear waves in nuclear media are 
analysed using  Nonlinear Schr\"odinger equation of fifth order and 
classified using a simple mechanical picture.
It is shown that a rich spectrum of nonlinear oscillations
in one-dimensional nuclear medium exist.
\end{abstract}

\end{frontmatter}
\newpage
\section{Introduction}
Nonlinear aspects of clusterization, one of the most mysterius 
and important problem of modern nuclear physics,
are of interest due to the following reasons.

First, practically all nuclear processes which are related to
clustering lead to large reconstruction of an initial nuclear
system and this requires a development of new theoretical methods
to describe large amplitude collective motions.

Second, the existance of clusters is a very general phenomenon. 
There are cluster objects 
in subnuclear and macro physics. 
Very different
theoretical methods were developed in these fields. However,
there are only few basic physical ideas, and most of methods
deal with nonlinear partial differential equations.

Third, methods of nonlinear dynamics give us possibility to derive
for nuclear physics unexpected collective modes,
which can not be obtained by traditional  methods of perturbation
theory near some equilibrium state.

The alpha decay is the most studied process
among the other possible
channels of fragmentations at low energies, such as
cluster radioactivity, cold fission and fusion\cite{SG89}.
As a rule these processes lead to a large mass transfer
up to the final channel. The existence of a central region 
of a constant density and a well shaped 
surface region makes it possible to describe a clusterization in
the region of nuclear surface 
including the neck region\cite{kar93pn}. 

The alpha decay and cluster radioactivity
indicate well the formation of a cluster on the surface 
of a large nucleus\cite{SG89}.
Our nonlinear approach was formulated 
in such a way to
describe the preformation
of such cluster\cite{LSG92ijmpe}.
Our model is
an extension of the Bohr-Mottelson
collective model\cite{EisGre}, 
which allows not only the nuclear deformations
which lead to collective vibrational and rotational motions, but
also the creation of bumps on nuclear surface.

Nuclear density falls considerably down in 
a region of nuclear surface. This fact is very important due
to possibility of a clusterization and other fluctuations
of a nuclear density leading to instability in 
a surface or a neck regions\cite{LSG92ijmpe},\cite{K93jpg}.
Recently\cite{KLSG96ijmpe} we suggested 
an alternative nonlinear approach,
based on the currents and density algebra\cite{kar84}
to describe a surface interaction of
a nuclei with a cluster.  
Starting from the hydrodinamic set of
equations and using an effective Skyrme interaction, 
a nonlinear Schr\"odinger equation with fifth order
term in the nonlinearity ($NoSE3-5$) 
has been deduced to describe in
a semiclassical limit an irrotational flux of 
a nuclear density. In the framework of this approach  
an approximate way to describe 
surface axial symmetric interactions of a large target with
a small cluster has been derived.

\begin{figure}[h,t,b]
\includegraphics[angle=-90,width=84mm]{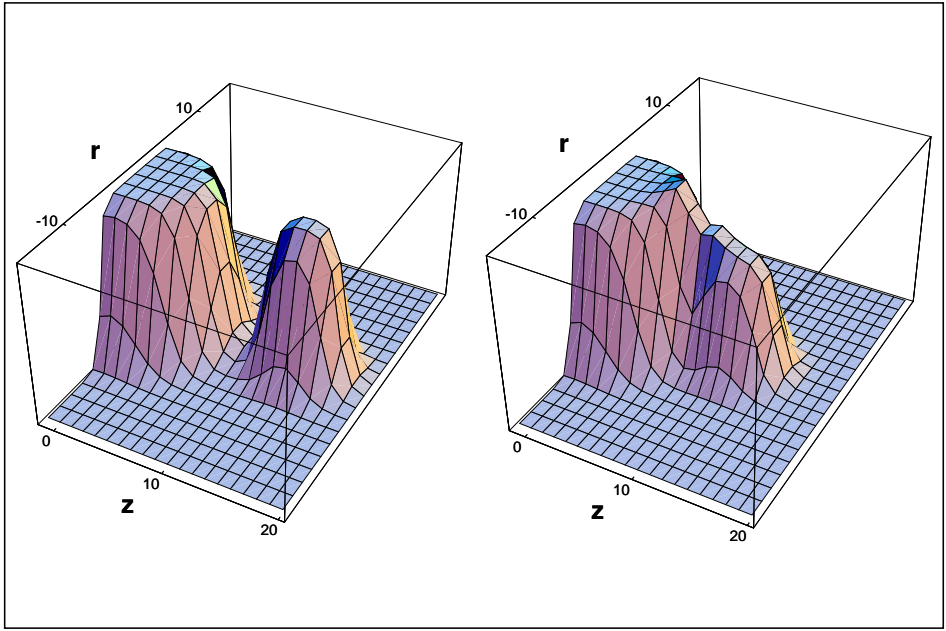}      
\vspace*{-15mm}
\caption{\small Surface interaction of $\pb$ and $\mg$, as a superposition of
nonlinear waves of NoSE3-5} 
\end{figure}
Figure~1 gives the density distributions for a surface
axial symmetric (along the $z$-axes) interaction
of a large target with a small cluster,
as  an interference of two 
nonlinear solutions of NoSE3-5. 
One can see the transition from well localized and separated 
initial distribution to a complicated picture, when 
the nucleus $\mg$ begins to be absorbed in the surface region. 

Methods of nonlinear dynamics are
rather unusual for traditional nuclear theory. 
Therefore it would be useful 
to understand: 
Why there exist nonlinear effects within linear
Schr\"odinger picture with a many-body nuclear hamiltonian? 
Does there exist a way to obtain nonlinear evolution equations, 
starting with traditional Hamilton or Lagrange picture?
How to classify possible types of solutions and to understand 
their physical meaning?

Within this paper we shall try to clarify these points
aimed to consider simplest
large amplitude oscillations of nuclear density.

The paper first 
presents in Sec.2 the basic scheme of our nonlinear approach to 
nuclear hydrodynamics. 
In Sec.3 it is shown how to reduce 
a set of hydrodynamical equations to
a dimensionless nonlinear Schr\"odinger equation.
Properties of nonlinear stationary waves are investigated in Sec.4.
We give the answers to these three above
questions and present a short summary in the last section .

\section{Nonlinear approach to nuclear hydrodynamics}
Fluid dynamical concept\cite{madelung} of the nuclear medium is a
traditional way to classify collective modes. Hydrodymamics gives up
natural set of variables(collective currents and density) to describe
any collective excitations of nuclear density. At present there are
a lot of formulations of the nuclear hydrodynamical approach. We will follow
the way, which is based on the current and density operators 
algebra\cite{kar93pn},\cite{kar84}

\subsection{General scheme}
For nuclear systems a usual second-quantized
form of a nonrelativistic theory is defined by introducing 
canonically conjugated nucleon fields
$\Psi ({\vec{x}},\vec{\sigma},q)$, where
$(q)$ denotes ($n,p$) for neutrons and protons and 
$\sigma = \pm$ the spin index,
which satisfy the equal-time
canonical anticommutation relations
\begin{equation}
\bigl\{ \Psi^{+}(1) , \Psi(2) \bigr\} _{+} = \delta(1-2),\qquad
\bigl\{ \Psi^{+}(1) , \Psi^{+}(2) \bigr\} _{+} = 
\bigl\{ \Psi(1) , \Psi(2) \bigr\} _{+} = 0.
\label{scheme1}
\end{equation}
In terms of these operators a nonrelativistic
Hamiltonian could be defined 
\begin{equation}
{\hat H}_{nucl} \bigl[ \Psi^{+}, \Psi;\; U_{nucl}(\vec{x}) \bigr]
\label{scheme2}
\end{equation}
with a general two-body nuclear interaction $U_{nucl}(\vec{x})$\\

The term "hydrodynamics" means that we will describe the dynamical
behavior of the nuclear system in a restricted space of collective variables
representing the density and the nucleon current of the system:
\begin{equation}
\Bigl( \Psi^{+}({\vec{x}}),\; \Psi({\vec{x}}) \bigr) 
\Longrightarrow
\Bigl(  {\hat \rho} ({\vec x}),\;  
{\hat j}_{k}({\vec x}) \Bigr) 
\label{scheme3}
\end{equation}
where the index $k$ denotes the cartesian vector components. 

The description of the evolution in terms of collective
currents (\ref{scheme3}) requires the of use commutation relations
between the collective current and density operators 
\begin{equation}
[ {\hat \rho} (\vec{x}) , {\hat j}_{k}(\vec{y}) ],\quad
[ {\hat \rho} ({\vec{x}}) , {\hat \rho} ({\vec{y}}) ],\quad
[ {\hat j}_{k}(\vec{x}) , {\hat j}_{l}({\vec{y}}) ],
\label{scheme4}
\end{equation}
and their commutation
relations with the hamiltonian (\ref{scheme2})
\begin{equation} 
{\partial {\hat \rho} (\vec{x})\over \partial t} = {1 \over i\hbar}
[ {\hat \rho} (\vec{x}) , {\hat H}_{nucl} ],\qquad 
{\partial {\hat j}_{k} (\vec{x}) \over \partial t} = {1 \over i\hbar}
[ {\hat j}_{k}(\vec{x}) , {\hat H}_{nucl} ]. 
\label{scheme5}
\end{equation}
Commutation relations (\ref{scheme4},\ref{scheme5}) 
can be obtained by using the
initial commutation relations (\ref{scheme1}) for nucleon
quantum fields and the existing formal representation 
for the kinetic energy density tensor operator
\begin{equation}
{\hat T}_{nk}({\vec{x}}) \equiv {\nabla_{n}}\Psi^{+}(
{\vec{x}})\cdot {\nabla_{k}}\Psi ({\vec{x}}) +
{\nabla_{k}}\Psi^{+}({\vec{x}})\cdot
{\nabla_{n}}\Psi ({\vec{x}})
\label{scheme6}
\end{equation}
in terms of collective currents
\begin{equation}
{\hat T}^{hyd}_{kl} = ( {\hat j}_{k}
{\hat \rho}^{-1} {\hat j}_{l} + {\hat j}_{l}
{\hat \rho}^{-1} {\hat j}_{k} ) +
{1 \over 2} {\hat \rho}^{-1}
{\nabla_{k}}{\hat \rho}\cdot {\nabla_{l}}{\hat \rho} 
+ \mbox{const} \cdot \delta_{kl}
\label{scheme7}
\end{equation}
which satisfies Eqs.~(\ref{scheme5}) as well.

As a result one gets a collective hydrodynamical
hamiltonian 
\begin{equation}
{\hat H}_{nucl} \bigl[ \Psi^{+};\; \Psi;\; U_{nucl}(\vec{x}) \bigr]
\Longrightarrow 
{\hat H}_{hyd} \bigl[ \rho;\;\vec{j};\;
T^{hyd},\;{\mathfrak E}[\rho,\vec{j}] \bigr]
\label{scheme8}
\end{equation}
which is equivalent to the initial nuclear hamiltonian
(\ref{scheme2})
relative to the equations of motion (\ref{scheme5}).
Here ${\mathfrak E}[\rho,\vec{j}]$ denotes effective potential term. 
We present the hamiltonian  only schematically aimed 
to show the general scheme.

It is convenient to use Skyrme interaction, 
which gives an ansatz for an effective hamiltonian 
${\hat H}_{hyd}  \Longrightarrow \int {\d}^{3}x \/{\hat H} ({\vec x})$
 of a local operator in terms of densities and currents\cite{}
\begin{eqnarray}\nonumber
{\hat H} ({\vec x}) &=& (\hbar^{2}/{2m}) \tau + B_{1}\rho^{2} +
B_{2}(\rho_{n}^{2}+\rho_{p}^{2})+B_{3}(\rho\tau-\vec{j}^{2})\\ \nonumber
&+&B_{4}(\rho_{n}\tau_{n}-\vec{j}_{n}^{2}+
\rho_{p}\tau_{p}-\vec{j}_{p}^{2})+B_{5}\rho\Delta\rho+
B_{6}(\rho_{n}\Delta\rho_{n}+\rho_{p}\Delta\rho_{p})\\ \nonumber
&+&B_{7}\rho^{2+\alpha}
+B_{8}\rho^{\alpha}(\rho_{n}^{2}+\rho_{p}^{2}) +
B_{9}(\rho\nabla\cdot\vec{\Xi}+\vec{j}\cdot\nabla\times\vec{\Sigma}\cr
&+&
\rho_{n}\nabla\cdot\vec{\Xi}_{n}+\vec{j}_{n}\cdot\nabla\times\vec{\Sigma}_{n}
+\rho_{p}\nabla\cdot\vec{\Xi}_{p}+
\vec{j}_{p}\cdot\nabla\times\vec{\Sigma}_{p})\\ 
&+&B_{10}\vec{\Sigma}^{2}+B_{11}(\vec{\Sigma}_{n}^{2}+\vec{\Sigma}_{p}^{2})
+B_{12}\rho^{\alpha}\vec{\Sigma}^{2}+B_{13}\rho^{\alpha}
(\vec{\Sigma}_{n}^{2}+\vec{\Sigma}_{p}^{2}),
\label{ham1}
\end{eqnarray}
where the numerical coefficients $B_{i}$ and $\alpha$ are parameters
of a Skyrme interaction\cite{bfh87}.

To describe a system of nonrelativistic spin-1/2 nucleons in terms
of currents requires 
to decompose spin-isospin density tensors  
to spin-scalar
\begin{equation}
\rho_{q}(\vec{x}) \equiv \sum_{\sigma} \Psi^{+} ({\vec{x}},\sigma,q)
\Psi (\vec{x},\sigma,q)
\label{s1}
\end{equation}
and spin-vector densities
\begin{equation}
\vec{\Sigma}_{q}(\vec{x}) \equiv \sum_{\sigma,\sigma'}
\Psi^{+} (\vec{x},\sigma,q)
<\sigma |\vec\sigma|\sigma'> \Psi (\vec{x},\sigma',q).
\label{s2}
\end{equation}
The related currents are the spin-scalar 
\begin{equation}
j_{k}({\vec x})_{q} = {1\over 2i} \sum_{\sigma}\Bigl( \Psi^{+}(
{\vec{x}},\sigma,q) {\nabla_{k}}
\Psi ({\vec{x}},\sigma,q) - {\nabla_{k}}
\Psi^{+}({\vec{x}},\sigma,q)\cdot\Psi ({\vec{x},\sigma,q}) \Bigr),
\label{s3}
\end{equation}
and spin-orbital ones
\begin{eqnarray}
\Xi_{k}({\vec x})_{q} &=& {1\over 2i} \sum_{\sigma,\sigma'}
\sum_{m,n=1}^{3} \varepsilon_{kmn}\Bigl( \Psi^{+}(
{\vec{x}},\sigma,q) {\nabla_{m}}
\Psi ({\vec{x}},\sigma',q) \\ \nonumber
&-& {\nabla_{m}}
\Psi^{+}({\vec{x}},\sigma,q)\Psi ({\vec{x},\sigma',q}) \Bigr)
<\sigma|\sigma_{n}|\sigma'>.
\label{s4}
\end{eqnarray}

Among all components of the kinetic-energy-spin-tensor the most
important ones are the scalars in spin and in the ordinary coordinate space
\begin{equation}
\tau_{q},\quad \vec{j}_{q}^{2},\quad \vec{\Sigma}_{q}^{2},\quad
\nabla\cdot\vec{\Xi}_{q},
\end{equation}
\begin{equation}
\tau =  \frac{1}{2} \sum_{k} T_{kk} =
 \sum_{k}{\hat j}_{k}
{\hat \rho}^{-1} {\hat j}_{k} +
{1 \over 4} {\hat \rho}^{-1}
{\nabla_{k}} {\hat \rho} {\nabla_{k}} {\hat \rho}.
\label{s5}
\end{equation}

It should be noted that the operators (\ref{s4}-\ref{s5})
is not the complete set of 
spin-isospin tensor components of
the current-spin-tensor and
kinetic-energy-spin-tensor.
The detail analysis of the equation of motion (\ref{scheme5})
for all defined densities and currents could help to clarify the situation.
First of all it is necessary to define the effective hamiltonian.

In order to analyze the evolution equations (\ref{scheme5}) 
with the effective
hydrodynamical hamiltonian (\ref{ham1}) it is necessary: first
to derive commutation relations between all
components of spin-isospin current and density tensors, and their 
commutation relations of with all terms of hamiltonian (\ref{ham1})
and second
 to provide tensor decomposition of all currents to 
irreducible cartesian components aimed to separate irrotational and
rotational terms.

This task is rather difficult. However,
in the framework of presented above general quantum hydrodynamics,
one could build, in principle, 
a quantum approach to investigate spin-scalar,
spin-vector, spin-orbital vibrational and rotational modes 
in nuclei. Such investigations are in progress.

\section{Nonlinear Schr\"odinger equation}
In order to check the possibilities of our approach
we simplify 
the hamiltonian (\ref{ham1}) aimed  
to describe different
oscillations of isoscalar nuclear density.
We assume a
system of spinless and isospinless nucleons
with no Coulomb, spin and spin-orbital effects ($W_{0}=0$)
and effective Skyrme forces (\ref{ham1}) ($\alpha = 1$)
\begin{eqnarray}\nonumber
H ({\vec x}) &=&
\frac{\hbar^{2}}{2m} \tau + \frac{3}{8} t_{0}\rho^{2} +
\frac{1}{16} t_{3} \rho^{3} +
)+ \frac{1}{64}(9 t_{1}-5t_{2}) |\nabla\rho |^{2} \\
&+& \frac{1}{8}(3 t_{1}+5t_{2}) (\{\rho,\tau\}_{+} -
2\vec{j}^{2}.
\label{hmod0}
\end{eqnarray}

Schematic "nonlinear density and current dependence" 
of kinetic energy is
\begin{equation}
H ({\vec x}) \Longrightarrow  H_{mod}({\vec x}) =
\frac{\hbar^{2}}{2m^{*}} \sum\limits_{k=1}^{3} j_{k}\frac{1}{\rho}j_{k}
+ \frac{\hbar ^{2}\eta^{2}}{8m^{*}}\frac{|\nabla\rho|^{2}}{\rho}
+ \frac{3}{8} t_{0}\rho^{2} +
\frac{1}{16} t_{3} \rho^{3} ,
\label{hmod1}
\end{equation}
where the following renormalization is used
\begin{eqnarray}\nonumber
m &\Longrightarrow& m^{*} \equiv (m^{-1} + (3t_{1} + 5t_{2})
\rho _{N}/8\hbar ^{2})^{-1},\\ 
{\hbar ^{2} \over 8m} &\Longrightarrow& {\hbar ^{2}\eta^{2} \over 8m} \equiv
{\hbar ^{2} \over 8m} + {\rho _{N} \over 64} (9t_{1} - 5t_{2}).
\label{hmod3}
\end{eqnarray}

$H_{mod}$ form Eq.~(\ref{hmod1}) is the simplest hamiltonian which includes
the "collective" kinetic energy,
"Weizs{\"a}cker" kinetic part which includes the
quantum and other gradient terms and the
"potential" energy, containing effective
two-body attractive term and three-body imitation of Pauli principle.\\

In classical hydrodynamics\cite{Lamb},
the fundamental variables are the local density
$\rho (\vec{x})$ 
and the fluid velocity
$\vec{v}(\vec{x})$ 
from which the  quantum current density can be defined by the 
anticommutator\cite{geilikman}\footnote{See discussion on a velocity
concept in quantum hydrodynamics in a recent paper\cite{KGMG96}}        
\begin{equation}
\hat{j}_{i}(\vec{x}) = {m \over 2\hbar} \{ {\hat \rho} (\vec{x}), 
\hat{v}_{i}(\vec{x}) \}_{+}.
\label{ve0}
\end{equation}

Using the definition (\ref{ve0}) and the commutative relations between
density and current operators (\ref{scheme4}) one derives 
velocity-density and velocity-velocity commutative relations
\begin{equation}
[ {\hat v}_{k}(\vec{x}), {\hat \rho} ({\vec{y}}) ],\qquad
[ {\hat \rho} ({\vec{x}}) , {\hat \rho} ({\vec{y}}) ],\qquad
[ {\hat v}_{i}(\vec{x}), {\hat v}_{j}(\vec{y}) ].
\label{ve1}
\end{equation}

To classify the type of motions in the cartesian coordinate space 
it would  be
convenient to provide tensor decomposition  of
the operator vector field
\begin{equation}
\hat{v}_{k}(\vec{x}) =
\nabla_{k}\hat{\Phi}_{pot}(\vec{x})  + curl_{k}(\hat{\vec{A}}),\qquad
div(\hat{\vec{A}}) = 0,
\label{helm1}
\end{equation}
into a sum of potential
$(\hat{\Phi}_{pot}(\vec{x}))$ and solenoidal
$(\hat{\vec{A}})$ fields
by using the Helmholtz theorem (see, e.g., ref.\cite{morsfesh} vol.~1).

In the following we shall
consider the equations of motion (\ref{scheme5}) with the hamiltonian
$H_{mod}$ (\ref{hmod1}). First
it is convenient to use the scale 
transformations~\cite{kar84} 
\begin{equation}
\vec{x} \equiv ( \hbar \eta / m^{*} c_{s} ) \/ \vec{x}^{'}, \qquad
t \equiv (m^{*} c_{s}^{2} / \hbar) \/ \tau,
\label{scale1}
\end{equation}
\begin{equation}
v_{k} (\vec {x}) = c_{s} U_{k}(\vec{x}'),\qquad 
\rho (\vec{x}, t) \equiv \rho _{N} \/ n ^{2} (\vec{x}', \tau).
\label{scale2}
\end{equation}
\noindent where $ c_{s} $ is the sound velocity in nuclear matter
\begin{equation}
c_{s} \equiv \bigl( {1 \over m^{*}} \rho ^{2} {\delta ^{2}
({\mathfrak E} / \rho ) \over \delta \rho ^{2} } \vert _{\rho =\rho_{N}} \bigr)
^{1/2}
\label{scale3}
\end{equation}
The scale transformation (\ref{scale1},\ref{scale2}) makes it possible
to separate the general effects related to
nonlinearity in the hamiltonian (\ref{hmod1}) (a polinomial in
density) from effects related to a choice of parameters of interaction,
which define only scales factors: effective length wave 
($\hbar \eta / m^{*} c_{s}$ ), the density of nuclear matter $\rho_{N}$,
the sound velocity in nuclear matter $c_{s}$ and the time factor
($m^{*} c_{s}^{2} / \hbar$).  

Hamiltonian (\ref{hmod1}) can be cast in the following form
\begin{equation}
H_{mod} = \rho_{N} \left(\frac{\hbar\eta}{m^{*}c_{s}}\right)^{3}
m^{*}c_{s}^{2}
\int {\d}^{3}x' \left( \sum\limits_{k=1}^{3} \frac{U_{k} n U_{k}}{2}
+ \frac{|\nabla n|^{2}}{8n} - n^{2} +
\frac{1}{2}n^{3} \right).
\label{hmod2}
\end{equation}
Now the continuity and Euler operator equations (\ref{scheme5})
can be reduced to the following set of  
the  dimensionless equations of motion
\begin{equation}
\frac{\partial n (\vec {x}')}{\partial\tau} = -\frac{1}{2}
\sum\limits_{k=1}^{3} 
\frac{\partial}{\partial x_{k}'} \{ {n (\vec {x}'), 
U_{k} (\vec {x}')}\}_{+},
\label{scale4}
\end{equation}
\begin{eqnarray}\nonumber
\frac{\partial U_{k} (\vec {x}')}{\partial\tau}  &=& \frac{1}{2}\left(
\left( \vec {U}\times curl \vec {U}\right)_{k} - 
\left( curl \vec {U}\times \vec {U}\right)_{k}\right)\\ 
 &-& \frac{1}{2}
\frac{\partial}{\partial x_{k}'}\left( \vec {U}^{2} +
\frac{\Delta n}{2n} - \frac{|\nabla n |^{2}}{4n^{2}}
- 4 n (\vec {x}') + 3 n^{2} (\vec {x}') \right).
\label{scale5}
\end{eqnarray}

Taking into account the velocity-velocity and 
velocity-density commutative relations (\ref{ve0}-\ref{ve1}) 
and the operator Helmholtz theorem (\ref{helm1}) we obtain
the following decomposition of the dimensionless velocity operator:
\begin{equation}
{\hat U}_{k}(\vec{x}) = \frac{\partial{\hat\Phi}(\vec{x})}{\partial x_{k}}
+ {\hat R}_{k}(\vec{x}),\qquad div\vec{R}=0,\qquad 
\vec{\hat\zeta} = curl\vec{\hat{R}},
\label{scale6}
\end{equation}
into potential and pure rotational currents.

Considering only the vibrational type of motion and neglecting
their possible
connection with rotational modes ($\vec{\hat\zeta}=0,\, {\hat R}_{k} \to 0$)
the commutation relations between density and potential velocity  are reduced
to the canonical boson form
\begin{equation}
[ \hat{n}(\vec{x}), \hat\Phi(\vec{y}) ] = i \delta(\vec{x}-\vec{y}),\qquad
[ \hat\Phi(\vec{x}), \hat\Phi(\vec{y}) ] = 0,\qquad
[ {\hat n} ({\vec{x}}) , {\hat n} ({\vec{y}}) ] = 0,
\label{scale7}
\end{equation}
which give it possible to treat (if it is no  rotation ($ R_{k}=0 $))
the evolution via $ H_{mod} $ as nonlinear operator anharmonical picture.

In the semiclassical limit the two hydrodynamical equations of motion
(\ref{scale4},\ref{scale5})
in the case of irrotational flow 
(\ref{scale6},\ref{scale7})
 can be cast in the form of 
one nonlinear Schr\"odinger equation 
\begin{equation}
i {\psi_\tau} = -\Delta
\psi - 4 \mid \psi \mid ^{2} \/ \psi +
3 \mid \psi \mid ^{4} \/ \psi
\label{nls00}
\end{equation}
for a complex function $\psi$ 
\begin{equation}
n(\vec{x},\tau) = \mid \psi(\vec{x},\tau) \mid ^{2},\qquad
\Phi(\vec{x},\tau) = \arg \psi(\vec{x},\tau).
\label{nls000}
\end{equation}
Eqs.~(\ref{nls00},\ref{nls000}) become transparent if we recall
the well-known formal analogy of quantum mechanics to 
fluid mechanics\cite{madelung}. However one should keep in mind
that in formulaes (\ref{nls00},\ref{nls000}) all variables are
collective ones and the effective potential is defined by a first
functional derivative of effective interaction on density.

\section{The one-dimensional nonlinear waves}
Let us consider the
stationary (travelling) solutions 
of the one-dimension (in the cartesian coordinate
space) 
nonlinear Schr{\"o}dinger equation (\ref{nls00}).
These solutions take the usual form
\begin{equation}
\psi(r, \tau) \equiv \psi(x)
\exp \bigl( i(\lambda - v^{2} / 2) \tau + i xv + \sigma _{0})
\label{nls02}
\end{equation}
where $r = x - v\tau - r_{0}$.
The parameters  $r_{0}, \sigma _{0}, v$
are the initial position of the center of the wave, 
the initial phase shift and the initial velocity.

The function  $\psi(x)$  is the solution of the nonlinear 
Schr\"odinger equation
\begin{equation}
\psi_{xx} - \lambda \psi + 4  \psi^{3}
- 3  \psi^{5} = 0.
\label{nls04}
\end{equation}


\begin{figure}[h,t,b]
\vspace*{15mm}
\includegraphics[width=90mm]{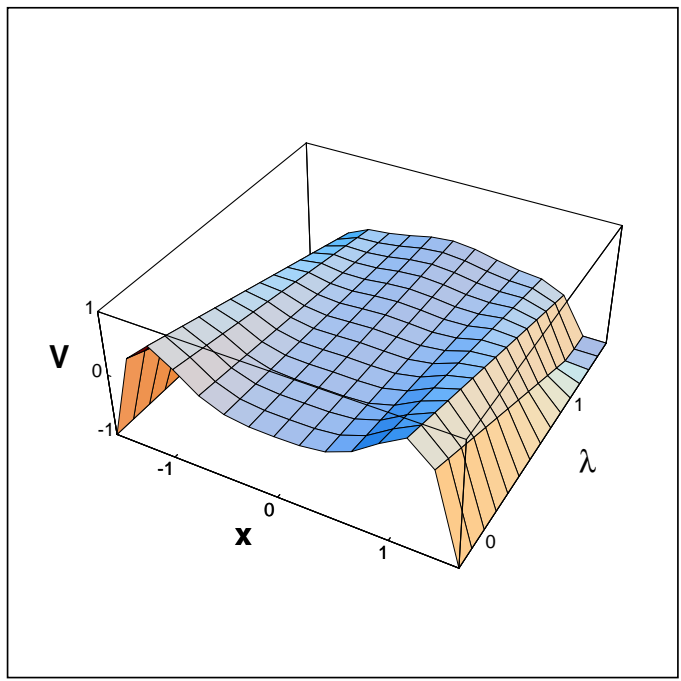}      
\caption{\small Dependence of "potential" $V(\psi)$ on $\psi$ and
and $\lambda$}
\vspace*{3mm}
\end{figure}

To analyze different solutions of Eq.~(\ref{nls04}) 
we will
exploit its simple mechanical analog, in which 
a macroscopic point "particle" of unit mass is moving
under a conservative "force"
\begin{equation}
F(\psi) = \lambda \psi - 4  \psi^{3} + 3  \psi^{5} \equiv
- \frac{\partial V(\psi)}{\partial\psi},
\label{nls05}
\end{equation}
where the "potential" (see Figure~2) is taken as
\begin{equation}
V(\psi) = -\frac{\lambda}{2}\psi^{2} + \psi^{4} - \frac{1}{2}\psi^{6},
\label{nls06}
\end{equation}
with $\psi$ as a "particle position"
and $x$ its "time".

Eq.~(\ref{nls04}) after being multipled by $\psi_{x}$ ,
can be one time integrated. The result is
\begin{equation}
\psi_{x} = \pm \sqrt{2(E - V(\psi))},\qquad
E = V(\psi) + \frac{1}{2} \psi_{x}^2
\label{nls07}
\end{equation}
where the constant of integration ($E$) is
considered to be the  "energy".
Eq.~(\ref{nls07}) is just the energy conservation law in the
analog "particle" problem.
The parameter ($\lambda$) defines the "potential" profile 
and the "energy"
( $E$ ) defines the "particle" motion.

Figure~2 gives a gross "potential" picture.
We show a detail set of "potentials" $V(\psi)$ for different
values of the parameter ($\lambda$ ) in Figs.~3 .
The "potential" $V(\psi)$ (\ref{nls06}) vanishes at the points
\begin{equation}
\psi = \{ 0,\quad \pm P_{1}=\pm\sqrt{1-\sqrt{1-\lambda}},\quad
\pm P_{2}=\pm\sqrt{1+\sqrt{1-\lambda}} \}
\label{nls08}
\end{equation}
(see Figure~3~($\lambda=0.5$)).

The equilibrium points of motion $F(\psi) = 
-\partial V(\psi)/\partial\psi = 0$
are given by
\begin{equation}
\psi = \{ 0,\quad \pm P_{3}=\pm\sqrt{\frac{2-\sqrt{4-3\lambda}}{3}},
\quad   
\pm P_{4}=\pm\sqrt{\frac{2+\sqrt{4-3\lambda}}{3}} \}
\label{nls09}
\end{equation}
(see Figs.3~($\lambda=0.9$)).
For an nonpositive $\lambda \leq 0$ 
a finite amplitude motion is possible only for
 $ 0 < E < V(P_{4}) $ and $ \vert \psi(0)\vert < P_{4} $ 
(see Figs.3~($\lambda=0.0$)).
For $ 0 < \lambda < 1$ there exist a finite amplitude motion 
eith different types for the three regions of the "energy"
$\{ V(P_{3}) < E < 0, \; E=0, \; 0 < E < V(P_{4})\}$
(see Figs.3~($\lambda=0.5,0.9$)). This is the most interesting case 
and in the following we will consider
it in detail.
For $ \lambda = 1$ the "barriers" 
$ V(-P_{4}) = V(0) = V(P_{4}) = 0 $ (Figs.3~($\lambda=1.0$)).
A finite amplitude motion for  $ 1 \leq \lambda < 4/3$ 
is possible only for the negative "energy"  
$  V(P_{3}) < E < 0 $ (Figs.3~($\lambda=1.1$)).
For $ \lambda > 4/3$ (see Figs.~3~($\lambda=1.5$))
any motion of a finite amplitude
is impossible. A "particle" comes away from the origin for any energy.

Eq.~(\ref{nls04}) is an ordinary differential equation of a second order.
For any set of initial values $\psi(x_{0}),\;\psi_{x}(x_{0})$ at initial
"time" $x_{0}$
one can obtain, in principle, the solution  of Eq.~(\ref{nls04})
numerically. 
We will
restrict ourselves 
only to stationary waves of a finite amplitude.
This means that for a given $ \lambda $  the "energy" is less than
the maximum "barrier" value ( $ E < V(\pm P_{4})) $.
Therefore it is convenient to select an "initial point" $x_{0}$, as
the point at "potential" profile in which 
\begin{equation}
 E = V(\psi_{0}), \qquad \psi_{x}(x_{0}) = 0,\qquad \psi_{0} \equiv
\psi(x_{0}).
\label{nls16}
\end{equation}
Because of the "energy conservation" the $E$ parameter is related
to the wave amplitude $\psi(0)$.
\begin{figure}[htbp]
\vspace*{10mm}
\hspace*{-13mm}
\includegraphics[width=135mm]{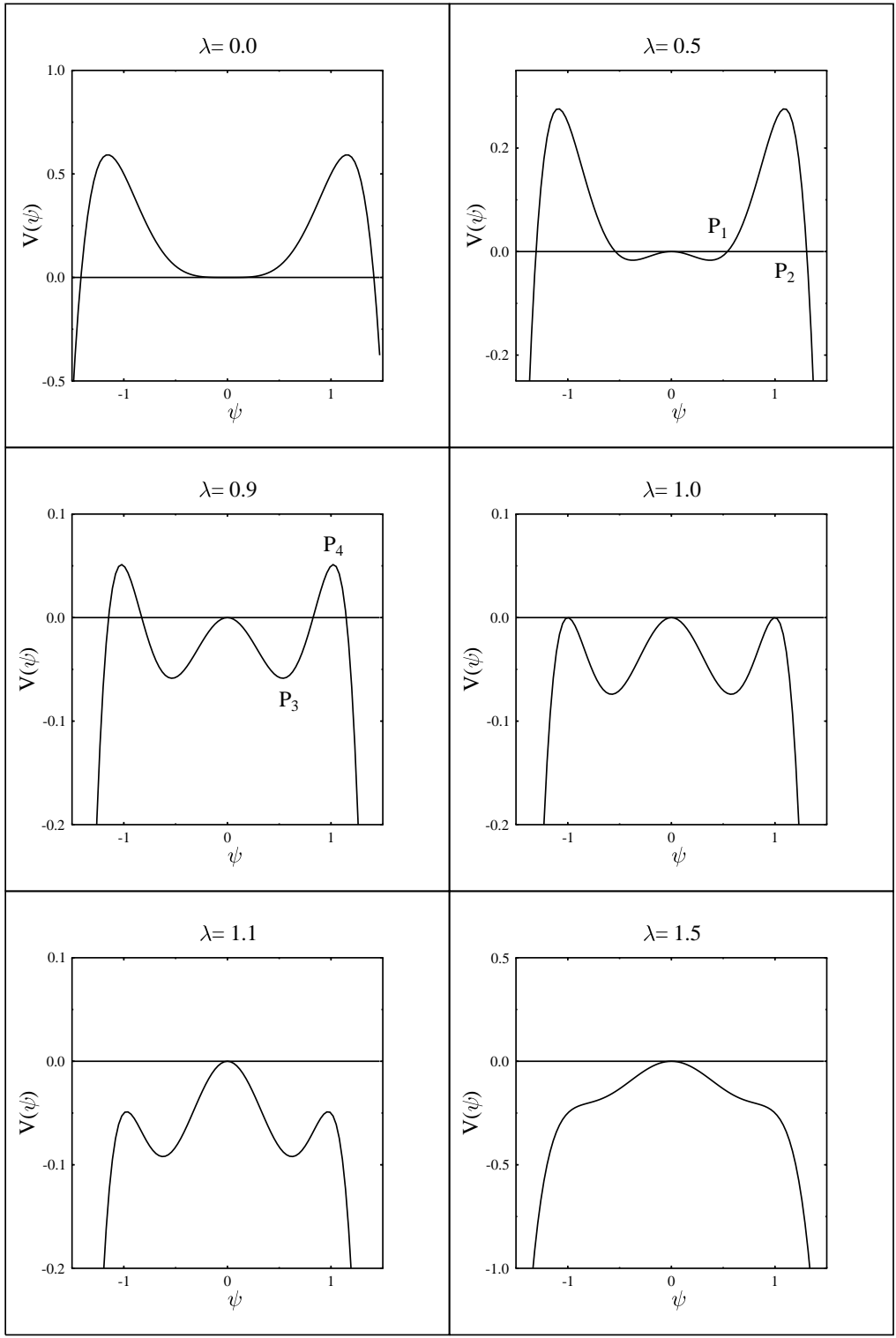}     
\caption{\small Dependence of a "potential" $V(\psi)$ on $\psi$
for different parameters $\lambda$}
\end{figure}
A localized soliton like solution
is the nodeless solution of Eq.~(\ref{nls04})  
under the following boundary condition
\begin{equation}
\psi(\infty) = \psi_{x}(\infty) = \psi_{x}(0) = 0.
\label{nls10}
\end{equation}
This corresponds to a special case with $ E=0 $. 
Soliton like solution exists if $ 0<\lambda<1$.  
A "particle" starts motion in the point $P_{1}$ or $-P_{1}$ 
(see Eq.~(\ref{nls08})) and comes to origin.

\begin{figure}[htbp]
\hspace*{-7mm}
\includegraphics[angle=-90,width=85mm]{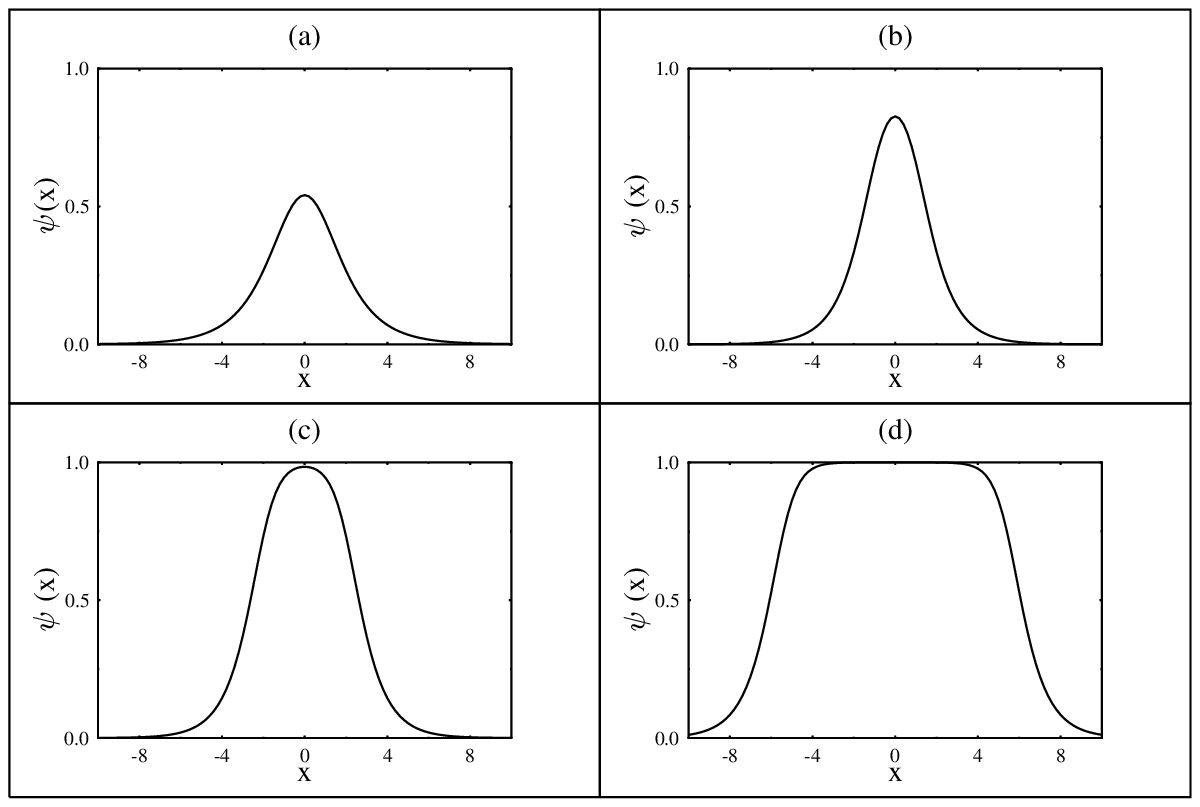}     
\vspace*{-18mm}
\caption{\small Soliton -like solutions of NoSE3-5 for 
different values of the parameters $\lambda$}
\end{figure}

In Figures~4 we present the soliton like solutions for a few values 
of the parameter ($\lambda$). 
There exist an analytical form of this soliton like solution\cite{kar84}
\begin{equation}
\psi(x) = (\lambda/(1 + (1-\lambda)^{1/2}\cosh(2\lambda^{1/2}r))^{1/2}.
\label{nls11}
\end{equation}
which looks like a symmetrized Fermi-distribution
\begin{equation}
\psi(x) =\sqrt{\frac{ 
\sinh ({\mathfrak R}/{\mathfrak D})}{\cosh ({\mathfrak R}/{\mathfrak D}) 
+ \cosh (x/{\mathfrak D})}}
\label{nls12}
\end{equation}
where the "radius"
\begin{equation}
{\mathfrak R} = \frac{1}{2\sqrt{\lambda}} {\rm arctanh} (\sqrt{\lambda}).
\label{nls13}
\end{equation}
and the "diffuseness"
\begin{equation}
{\mathfrak D} = 
\frac{1}{2\sqrt{\lambda}}
\label{nls14}
\end{equation}
are defined by the parameter ($\lambda$).\\
The parameter $\lambda$ is the chemical potential 
defined by the "particle number"
\begin{equation}
N = \int_{-\infty}^{\infty} \d x \; \psi(x)^{2},\qquad
N = {\rm arctanh} (\sqrt{\lambda})
\label{nls03}
\end{equation}

\begin{figure}[htbp]
\vspace*{35mm}
\hspace*{-10mm}
\includegraphics[width=135mm]{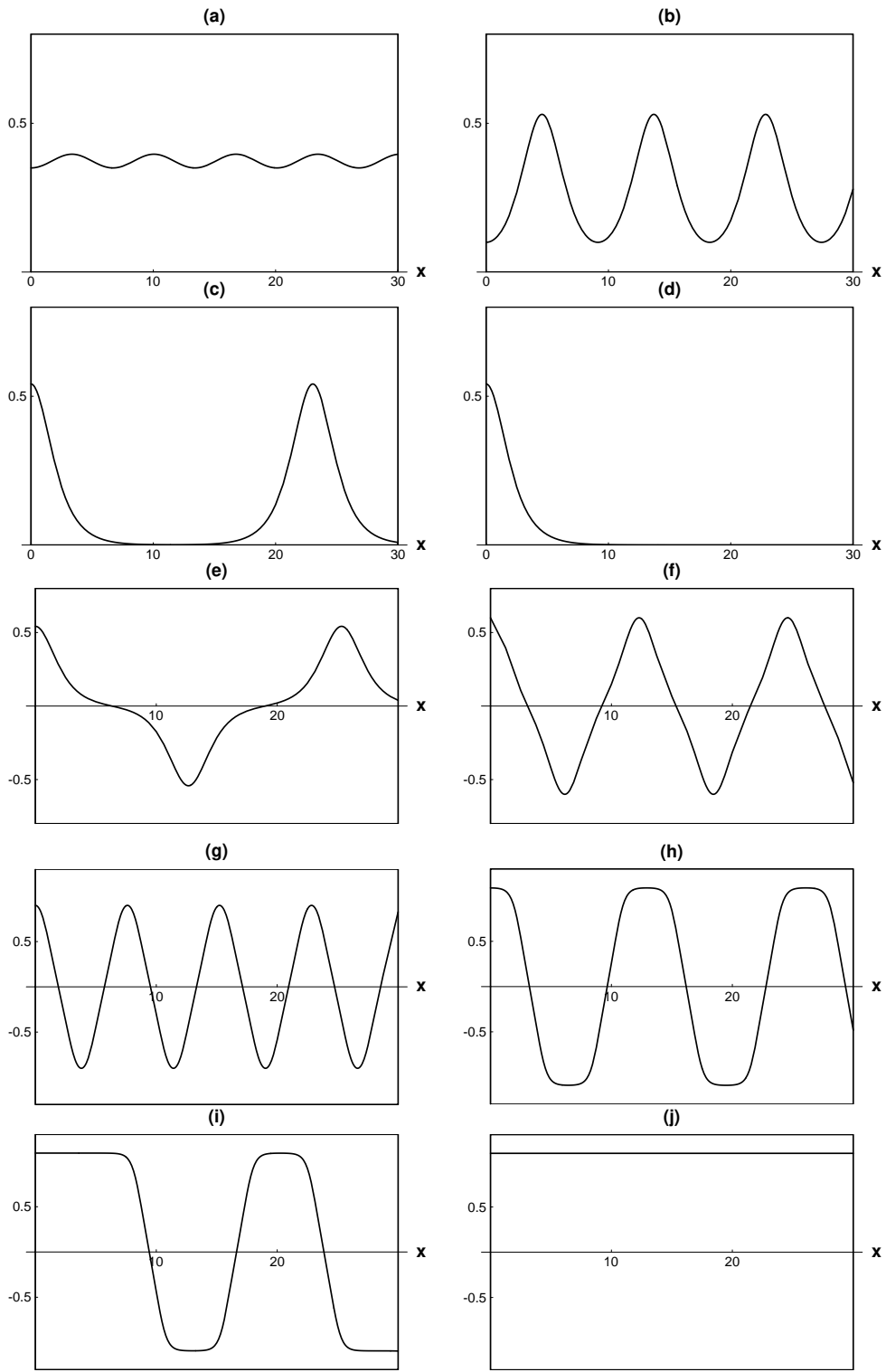}     
\vspace*{-20mm}
\caption{\small Set of stationary nonlinear waves of
 NoSE3-5}
\vspace*{-10mm}
\end{figure}

It is useful to note also the following features of this soliton like
solution. First for a small "particle number" the solution looks 
like a solution
of a cubic nonlinear Schr\"odinger equation.
Second, for an intermediate and large "particle numbers" 
 the "radius" (\ref{nls13}) is increased (see Figs.4).
Third,  there exist
an internal region of approximate constant "density" and 
a "surface region" of a "constant" diffuseness 
(see Eqs.~(\ref{nls13},\ref{nls14}) and Figs.4(c,d)).

In Figures~5 we present possible periodical and aperiodical solutions
of Eq.~(\ref{nls04}) for $ \lambda = 0.5 $. 
The corresponding "potential" is represented in Figs.3~($\lambda=0.5$).
One can see small oscillations with a positive amplitude
in the "well" for ($V(P_{3})  < E < 0$) (Figs.5(a))
These oscillations became  distorted more and more, as 
$ E \to 0-\epsilon$, where $\epsilon$ is an infinite small
positive value (see Figs.5(b,c)).
For $E = 0$ we have the exact soliton like solution (Figs.5(d)).
There are nonlinear waves of positive and negative apmlitude, as
$ E \to V(P_{3})+\epsilon$, (see Figs.5(e,f)).
These waves look like quasi-trigonometrical functions (see
Figs.5(g)), as $\lambda$ changes from small positive values to
the "barrier" $V(P_{4})$.

Periodical waves  became distorted more and more, as
$ E \to V(P_{4})-\epsilon$. 
In Figs.5(h,i) one can see a transition from
a well-defined periodical structure to an aperiodical one.

There is a region of semi "shock" wave with 
the well  defined sharp "front" (see Figs.5(i)). 
"Front" moves to the infinity
and, in this case, we have the constant solution in coordinate space 
when an "energy" is equal to the "barrier" ($E = V(P_{4})$) 
(see Fig.5(j)), describes a one-dimensional nuclear matter.  

The above solutions were obtained numerically.
However it is possible to get also analytical solutions 
for few special cases.
First it shoud be noted that it is possible to write a formal solution
of Eq.~(\ref{nls07})
\begin{equation}
\int_{\psi_{0}} \frac{{\psi} \,\d\psi}{\sqrt{2(E - V(\psi))}}= \pm 
\int_{x_{0}}{x} \d x
\label{nls17}
\end{equation}
for any types of nonlinearity, i.e. for any "potential" $V(\psi)$.

In our case $V(\psi)$ is given by a symmetrical
polinomial of six order in $\psi$ (\ref{nls06}).   
The substitution $Z \equiv \psi^{2}$ reduces 
equation (\ref{nls04}) to the following form
\begin{eqnarray}\nonumber
Z_{x} &=& \pm 2 \sqrt{Z (2E + \lambda Z - 2Z^{2} + Z^{3})}\\
 &\equiv& \pm 2 \sqrt{Z (Z-Z_{1}) (Z-Z_{2}) (Z-Z_{3})} 
\label{nls18}
\end{eqnarray}
The expression under the square root of the
Eq.~(\ref{nls18}) is a polinomial of fourth degree.
One of the roots is $Z=0$ the other three ones being defined by 
$E,\; \lambda $ parameters. 

The following solution of Eq.~(\ref{nls18}) exists also 
in the case of $ E=0, \lambda=0$
\begin{equation}
\frac{\sqrt{Z (Z - 2)}}{2Z} = \pm 2x\qquad Z(x) = \frac{2}{1-4x^{2}} 
\label{nls19}
\end{equation}
where the constants of integration are omitted. This solution has a pole
at $x = \pm 2$, and it does not describe a finite solution.

For very few special cases 
Eq.~(\ref{nls18}) can be integrated analytically 
in Jacobian elliptic 
functions.
These functions are quite similar to the trigonometric
functions. 
One can find in the book\cite{abramstegun} 
the complete set of twelve different
Jacobian ellipic functions and the coresponding form of the differential 
equation.
If it is possible 
to reduce Eq.~(\ref{nls18}) by a linear transformation 
$(Z, x) \to (z, u)$ to the one of these forms, then
the solution of (\ref{nls18})  can be cast as  
the related Jacobi elliptic function.
The explicit form of the solution is determined by the constants
$\lambda, E$ and boundary conditions. 

\section{Summary}
We give the following answers to the three questions
formulated in Introduction.

Nonlinear problems exist within a linear Schr\"odinger picture
with a many-body
nuclear hamiltonian as the result of the reduced description
in a restricted space of collective degrees of freedom.
In Sections 2 and 3 the way to derive the Nonlinear Schr\"odinger
equation for a collective motion is presented. 
The logic of our approach corresponds to the logic
of any microscopic or semimicroscopic theory of collective 
nuclear motion, namely,
one first selects a space of certain collective variables (\ref{scheme3})
 and their commutation relations (\ref{scheme4}). Then
seeks an expresssion for the collective Hamiltonian (\ref{scheme8})
 in terms of
the chosen collective variables  to reproduce the commutation
relations (\ref{scheme5}) of the original many-particle Hamiltonian 
and the collective operators.
The  hydrodynamical collective hamiltonian
 (\ref{scheme8}) is equivalent to the initial Hamiltonian (\ref{scheme2})  
only with respect to 
the equations of motion (\ref{scheme5})
for the collective 
operators (\ref{scheme2}).
The present approach is the realization for nuclear physics
of the general problem how to formulate nonrelativistic quantum mechanics
in terms of collective current operators.
The currents (\ref{scheme3}) can be considered as fundamental
dynamical variables rather than  the underlying canonical fields 
(\ref{scheme1}).
The main emphasis of this type of theory is the algebraic structure of the
equal-time current-current commutation relations (\ref{scheme4}).
The motivation for initiating such a program was due to
the fact that most of the interesting physical quantities
can be expressed in terms of currents and densities.
It is important for nuclear physics because the density and velocity
distributions are usual variables to analyse the dynamics of
heavy ion collisions (see the recent 
papers\cite{KGMG96},\cite{LSG97jpg}
and the references therein).

In the present paper, 
possible types of stationary nonlinear waves
in nuclear media are analyzed 
in the framework of Nonlinear Schr\"odinger
equation of fifth order 
and classified using a simple mechanical
picture. 
It is shown that a rich spectrum of
nonlinear oscillations 
in one-dimensional isoscalar nuclear matter exist.

\end{document}